\documentclass[twocolumn,showpacs,preprintnumbers,amsmath,amssymb,pra]{revtex4}
\usepackage{txfonts}
\usepackage{graphicx}
\usepackage{dcolumn}
\usepackage{bm}
\begin{document}


\title{Analog of photon-assisted tunneling originated from dark Floquet state in periodically modulated waveguide
arrays}
\author{Liping Li$^{1}$}
\author{Xiaobing Luo$^{2}$}\email{xiaobingluo2013@aliyun.com}
\author{Xiaoxue Yang$^{1}$}
\author{Xin-You L\"{u}$^{1}$}\email{xinyoulu@gmail.com}
\author{Ying Wu$^{1}$} \email{yingwu2@126.com}
\affiliation{$^{1}$ Wuhan National Laboratory for Optoelectronics and
School of Physics, Huazhong University of Science and Technology,
Wuhan, 430074, P. R. China}
\affiliation{$^{2}$ Department of Physics, Jinggangshan University, Ji$'$an 343009, P. R. China}

\date{\today}

\begin{abstract}
We theoretically report an analog of photon-assisted tunneling (PAT)
in a periodically driven lattice array without a static biased
potential by studying a three-channel waveguide system. This analog
of PAT can be achieved by only periodically modulating the top
waveguide and adjusting the distance between the bottom and its
adjacent waveguide. It is numerically shown that the PAT resonances
also exist in the five-channel waveguide system and probably exist
in the waveguide arrays with other odd numbers of waveguides, but
they will become weak as the number of waveguides increases. With
origin different from traditional PAT, this type of PAT found in our
work  is closely linked to the existence of the zero-energy (dark)
Floquet states. It is readily observable under currently accessible
experimental conditions and may be useful for controlling light
propagation in waveguide arrays.
\end{abstract}
\maketitle
\section{introduction}

Controlling quantum tunneling and transport through a periodic
driving field has been a subject of intense studies in the last
decades, for its relevance to fundamental physics tests as well as
to great potential application in nanoscale
devices\cite{Hanggi,Platero}. Among the most intriguing aspects of
the subject, coherent destruction of tunneling (CDT)\cite{Grossmann}
and photon-assisted tunneling (PAT)\cite{Tien} represent two seminal
results. CDT is a resonant effect discovered in the pioneering work,
in which the coherent tunneling between states is almost completely
suppressed when the system parameters are carefully chosen at the
isolated degeneracy point of quasi-energies\cite{Grossmann}. It has
so far generated great interests and has recently been observed
experimentally in different physical systems\cite{Valle,Kierig}.
Recently, CDT has been found to occur over a wide range of system
parameters in odd-$N$-state systems where one state is periodically
driven with respect to others\cite{Luo}. Such extension of
destruction of tunneling to a finite parameter range, referred to as
dark CDT, is attributed to the existence of localized dark Floquet
state with zero quasi-energy\cite{Luo,Luo2}. As the dark CDT was
introduced in the high-frequency regime where the driving frequency
is larger than all energy scales of the system, it naturally leads
to certain interesting questions: whether the dark CDT and
associated dark Floquet state still exist in the non-high-frequency
regimes which are accessible by adjusting, for example, either the
coupling strength between the neighboring states or the driving
frequency? what new behaviors will emerge in the non-high-frequency
regimes?

Photon-assisted tunneling (PAT) refers to a phenomenon in which
tunneling contact disabled by a static tilt (dc bias potential) can
be restored when the system exchanges energy of an integer number of
photons with the oscillating field\cite{Holthaus}. The static tilt
(dc bias potential) leads to suppression of tunneling which is
related to localized Wannier-Stark states\cite{Korsch}. When a
multiple of the driving frequency of ac field matches the energy
difference between adjacent rungs of the Wannier-Stark ladder, the
system is able to absorb or emit photons with sufficient energy to
bridge the energy difference created by the dc bias potential,
through which tunneling is (partly) restored (PAT). So far, PAT has
been experimentally observed in Josephson junctions\cite{Shapiro},
coupled quantum dots\cite{Kouwenhoven,Kouwenhoven2}, semiconductor
superlattices \cite{Guimaraes,Keay} and Bose-Einstein condensates in
optical lattices\cite{Sias}.

\begin{figure}[htb]
\centering{\includegraphics[width=0.5\textwidth]{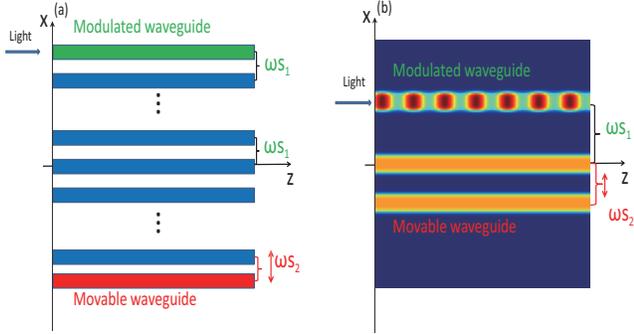}}
\caption{(Color online) (a) Schematic of a tunneling-coupled optical waveguide array
with controllable boundary that realizes an analog of
photon-assisted tunneling. (b) A typical triplet waveguide system.
The refractive index of the top boundary waveguide is modulated
periodically along the propagation direction. The space separation
between the bottom boundary waveguide and its nearest neighbor $
\omega s_2 $ is adjustable by moving the bottom boundary waveguide
towards the other waveguides, while the spacings $\omega s_1$
between other nearest-neighboring waveguides are fixed.}
\label{fig:fig1}
\end{figure}

In this article, we have studied the tunneling dynamics in lattice
arrays with controllable boundary. Owing to the simplicity and
flexibility offered by optical settings, the engineered photonic
waveguides provide an ideal system for exploration of tunneling
phenomena, in which spatial propagation of light mimics the temporal
dynamics of a quantum particle in a lattice
array\cite{Garanovich,Longhi}. As illustrated in Fig.\ref{fig:fig1},
we present an optical implementation of our Hamiltonian in the form
of a linear array of tunneling-coupled optical waveguides which is
characterized in: (i) that the refractive index of the top boundary
waveguide is modulated periodically along the propagation direction;
and (ii) that the distance $\omega s_2$ between the bottom boundary
waveguide and its nearest neighbor is different from other identical
nearest-neighboring spacings $\omega s_1$. Thus, through adjustment
of the distance $\omega s_2$, the coupling strength between the
bottom boundary waveguide and its nearest neighbor can be tuned to
be sufficiently large in comparison to the modulation frequency. The
role of photon is played by a periodic modulation of the the top
boundary waveguide with a certain modulation frequency. Generally,
PAT occurs in a system with a static biased potential which strongly
suppresses usual Josephson oscillations. But here we report an
analog of photon-assisted tunneling in a periodically driven lattice
array without static tilt (dc bias potential) by comprehensively studying a
three-channel waveguide system. Our numerical analysis discovers
that dark CDT (strong suppression of tunneling) and dark Floquet
state still exist in the three-channel waveguide system even in the
non-high-frequency regimes where the modulation frequency of the
periodically modulated top waveguide is roughly equal to or smaller
than the coupling strength between the bottom and its adjacent
(middle) waveguide. However, when integer multiples of the
modulation frequency approximately equal to the coupling strength
between the bottom and its adjacent (middle) waveguide, the light
tunneling from the top waveguide to the others is restored as a
clear signature of photon-assisted tunneling. Different from the PAT
observed in the earlier studies which usually requires a static
biased potential to initialize the system in a self-trapped state,
this type of PAT is closely linked to a dark Floquet state with zero
quasi-energy. Our results are applicable for the five-channel
waveguide system and also extendable to waveguide arrays with an odd
number of waveguides.

\section{Light tunneling dynamics in modulated photonic lattices with controllable boundary}
We now discuss light dynamics in a photonic structure composed of an
array of waveguides, in which the refractive index of the top
boundary waveguide is modulated periodically along the propagation
direction and the distance between the other boundary waveguide and
its nearest neighbor is adjustable, as schematically illustrated in
Fig.\ref{fig:fig1}. The coupled-mode equations describing the beam
dynamics in such structures can be written as

\begin{eqnarray}  \label{equ:PN} 
i\frac{{da_1 }}{{dz}}&=&\sigma(z)a_1 + \Omega _1 a_2, \nonumber \\
i\frac{{da_j }}{{dz}}&=&\Omega _1 a_{j-1} + \Omega _1 a_{j+1}, ~~~(j=2,3,...,N-2)\nonumber \\
i\frac{{da_{N-1}}}{{dz}}&=&\Omega _1 a_{N-2} + \Omega_2 a_N,  \nonumber \\
i\frac{{da_N }}{{dz}}&=&\Omega _2 a_{N-1},
\end{eqnarray}

where $ a_j $ is field mode amplitude in the  $j$-th waveguide, $z$
the propagation distance, $\Omega_1$ the coupling strength between
neighboring waveguides with spacing $\omega s_1$, $\Omega_2$ the
coupling strength between the bottom boundary waveguide ($j=N$) and
its adjacent waveguide ($j=N-1$), and $\sigma(z)$ the normalized
difference between the propagation constants of the top boundary
waveguide and the other waveguides of the array. We consider a
harmonic modulation of the linear refractive index of the top
boundary waveguide along the propagation direction with $\sigma(z)=A
\sin(\omega z) $, where $A$ is the relative depth of the harmonic
longitudinal modulation, and $\omega$ is the spatial modulation
frequency. In a different perspective, the above equation
(\ref{equ:PN}) can be regarded as describing the system of a quantum
wave in a periodically driven lattice array if $z$ is viewed as time
$t$.

As is well known, the periodic time-dependent equation
(\ref{equ:PN}) admits solutions in the form of Floquet states
$(a_1,a_2,...,a_N)^T=(\tilde{a}_1,\tilde{a}_2,...,\tilde{a}_N)^T\exp(-i\varepsilon
z)$, where $\varepsilon$ is  the quasi-energy and the  amplitudes
$(\tilde{a}_1,\tilde{a}_2,...,\tilde{a}_N)^T$ are periodic with
modulation period $T= 2\pi/\omega$. Our analysis is based on the
Floquet theory which offers a powerful tool for the treatment of the
periodically driven system.

In the case of $\Omega_1=\Omega_2 $, the equation (\ref{equ:PN}) is
reduced to the one discussed in Refs.~\cite{Luo, Luo2}, where a
significant suppression of tunneling (the so-called dark CDT) exists
for a wide range of system parameters in all odd-$N$-state systems
with identical coupling constants because of occurrence of the
zero-energy (dark) Floquet state. The case of $\Omega_1\neq\Omega_2$
has yet been explored. In this work, we will focus on that case and
give two example models (three- and five-waveguide systems) to
explore new physics in these systems.

\textbf{A. PAT in three-guide system}

We start our consideration for the three-guide system, the minimal
one for odd-$N$-state systems. In this case, the dynamical equations
are of the form
\begin{eqnarray}  \label{equ:P3}
i\frac{{da_{1}}}{{dz}} &=&A \sin(\omega z)a_{1}+\Omega _{1}a_{2},  \nonumber \\
i\frac{{da_{2}}}{{dz}} &=&\Omega _{1}a_{1}+\Omega _{2}a_{3},  \nonumber \\
i\frac{{da_{3}}}{{dz}} &=&\Omega _{2}a_{2}.
\end{eqnarray}

\begin{figure}[htb]
\centering{\includegraphics[width=0.48\textwidth] {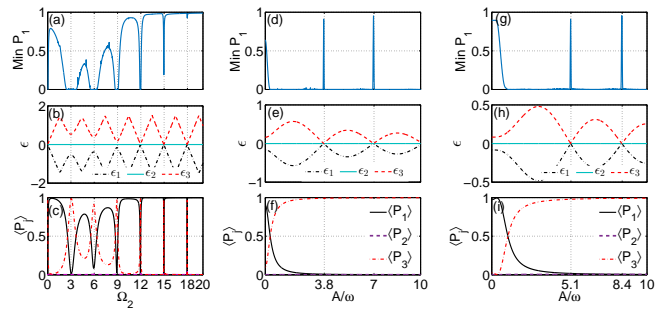}} \caption{(Color
online) PAT in three-guide optical system. The left column: (a) the
minimum value of intensity of light at the initially populated
guide-$1$, $ \textrm{Min} (P_1) $ versus $ \Omega_2$, with $A=6.6,
\omega=3, \Omega_1=1$; (b) the corresponding quasi-energy $ \epsilon
$ versus $ \Omega_2 $ and (c) the time-averaged population $ \langle
P_j \rangle $ belonging to the zero-energy (dark) Floquet state versus
$\Omega_2$. The middle column: (d) $ \textrm{Min} (P_1) $ versus $
A/\omega $ for the $1$-photon resonance $\omega=3, \Omega_2=3,
\Omega_1=1$; (e) the corresponding quasi-energy $ \epsilon $ versus
$ A/\omega $ and (f) the time-averaged population $ \langle P_j
\rangle $ belonging to the zero-energy (dark) Floquet state versus
$A/\omega $. The right column: (g) $\textrm{Min} (P_1)$ versus $
A/\omega $ for the $2$-photon resonance $\omega=3, \Omega_2=6,
\Omega_1=1$; (h) the corresponding quasi-energy $ \epsilon $ versus
$ A/\omega $ and (i) the time-averaged population $ \langle P_j
\rangle $ belonging to the zero-energy (dark) Floquet state versus $
A/\omega $.} \label{fig:fig2}
\end{figure}

To study the system's beam dynamics, we solve numerically the
coupled-mode equation (\ref{equ:P3}) with the light initially
localized in the $1$-th waveguide (the top boundary waveguide). With
the numerical solution, we compute the intensity of light staying in
the initial waveguide by $P_1(z)=|a_1(z)|^2$ and measure the minimum
value of $P_1(z)$ over a long-enough propagation distance. When
$\textrm{Min}(P_1)$ is not zero, the tunneling is suppressed as the
light is not allowed to be fully transferred from the $1$-th mode
(guide) to the other modes (guides). In Fig.\ref{fig:fig2}(a), we
display $\textrm{Min}(P_1)$ versus the coupling strength $ \Omega_2$
at the fixed parameters $ A=6.6 , \omega=3,  \Omega_1=1$. For $
\Omega_2=0$, the system is in fact a two-guide system in which the
conventional CDT happens only at the isolated degeneracy point of
the quasi-energies, and consequently $\textrm{Min}(P_1)$ takes a
zero value because our driving parameter is set to be slightly off
the isolated degeneracy point. When $\Omega_2$ is increased from
zero, the value of  $\textrm{Min}(P_1)$
 becomes relatively large except at a series of very sharp dips. In general, periodic modulation of the top boundary waveguide
 will yield a significant
 suppression of the light tunneling in the three-guide system even
 with $ \Omega_1\neq\Omega_2 $, as shown in Fig.\ref{fig:fig2}(a). However, at
particular values of the coupling strength $\Omega_2$,
$\Omega_2\approx
 n\omega$ with $n$ being integer, the value of  $\textrm{Min}(P_1)$ exhibits a series of sharp
 dips, in analogy to the $n$-photon-like resonances which destroy
 the effect of suppression of tunneling. It also can be observed
 that  as the coupling strength $\Omega_2$ is increased, the higher photon-like resonances become very weak and thus are almost not
 visible.

For a deep insight into the tunneling dynamics obtained in
Fig.\ref{fig:fig2}(a), we numerically compute the quasi-energies and
Floquet states of this system as shown in Fig.\ref{fig:fig2}(b)-(c).
As shown in Fig.\ref{fig:fig2}(b), this three-state system always
possesses a Floquet state with zero quasi-energy regardless of the
value of $\Omega_2$. The other two quasi-energies make a set of
close approaches to each other as $\Omega_2$ is increased. At the
points of close approach, namely, at $\Omega_2\approx
 n\omega$,
the value of  $\textrm{Min}(P_1)$ displays sharp dips and the
tunneling is significantly restored. We also plot the time-averaged
population distribution $\langle P_j \rangle=(\int_{0}^{T}dz
|a_j|^2)/T$ for the zero-energy Floquet state $(a_1,a_2,a_3)^T$ in
Fig.\ref{fig:fig2}(c). Considering that the dynamics is determined
by the Floquet state, self-trapping (suppression of tunneling) of
light intensity initially populating at the $1$-th mode (guide) will
take place if $\langle P_1 \rangle> 0.5$ holds. As seen in
Fig.\ref{fig:fig2}(c), the zero-energy Floquet state has negligible
population at the central mode (guide) while the population $\langle
P_ 1\rangle$ is much larger than $0.5$ for all values of $\Omega_2$
except those in the vicinity of $\Omega_2\approx n\omega$.
Correspondingly, suppression of tunneling (CDT) occurs for all
values of $\Omega_2$ except the locations of photon resonances, as
shown in Fig.\ref{fig:fig2}(a). The Floquet state with zero
quasi-energy is essentially the dark Floquet state, not only for its
zero quasi-energy but also for its negligible population at the
central waveguide; the suppression of tunneling (CDT) is of the dark
CDT as it is caused by the dark Floquet state rather than level
degenaracy. In fact, the CDT-PAT transition found in
Fig.\ref{fig:fig2}(b) is closely related to the sharp
localization-delocalization transition of population $\langle P_
1\rangle$ for the zero-energy (dark) Floquet state. Note that the
dark Floquet originally discovered and defined in the high-frequency
limit can be reduced to the well-known dark state by means of
high-frequency averaging method\cite{Luo, Luo2}. In this considered
model (\ref{equ:P3}), however, the dark Floquet state and the
associated CDT can still exist in the non-high-frequency regimes
where the coupling strength $\Omega_2$ is much larger than the
modulation frequency and the high-frequency averaging method is
invalid. These results will greatly enrich our understanding of dark
Floquet state and dark CDT.

In Fig.\ref{fig:fig2}(d), we show how the value of
$\textrm{Min}(P_1)$ varies under conditions that the modulation
amplitude is increased, while its frequency is held constant at
$\omega=3$ and the coupling strength held $\Omega_2=\omega,
\Omega_1=1$. This corresponds to the $n=1$ photon resonance. For
$A=0$, the system is self-trapped in the $1$-th waveguide due to the
existence of an imbalanced dark state $(-\Omega_2/\Omega_1,0,1)^T$
with $\Omega_2/\Omega_1>1$, and thus the value of
$\textrm{Min}(P_1)$ is nonzero. When the periodic driving is
applied, as $A$ is increased from zero, the value of
$\textrm{Min}(P_1)$ rapidly drops to zero, which indicates that the
photon resonance destroys the self-trapping effect. When $A/\omega$
is increased further, $\textrm{Min}(P_1)$ takes extremely low values
about zero except at a sequence of very narrow peaks. These peaks
are precisely centered at $A/\omega=3.83, 7.01$,..., the zeros of
$J_1(A/\omega)$. In Fig.\ref{fig:fig2}(e)-(f), we plot the
quasi-energies and the population distributions of the dark Floquet
state for the first photon resonance ($n=1$). Apparently, the
quasi-energies are degenerate when $J_1(A/\omega)=0$, and when away
from $A=0$ the dark Floquet state has averaged population at the
$1$-th mode (guide) well below the value of $0.5$. Therefore it can
be concluded that the well-defined quasienergy crossings instead of
the dark Floquet state are the origin of the extremely sharp peaks
seen in Fig.\ref{fig:fig2}(d).

In Fig.\ref{fig:fig2}(g), we show the values of $\textrm{Min}(P_1)$
in the three-guide optical system for the $n=2$ resonances. As is
expected, the values of $\textrm{Min}(P_1)$ exhibit a number of
extremely sharp peaks centered on the zeros of  $J_2(A/\omega)$
where the quasi-energies will be degenerate; see
Fig.\ref{fig:fig2}(h). Similar to the case of $n=1$ resonance, the
sharp peaks in the $n=2$ resonance is also caused by the level
degeneracy rather than the dark Floquet state, as it is shown in
Fig.\ref{fig:fig2}(i) that the population $ \langle P_1 \rangle $
corresponding to the dark Floquet state is well below $0.5$ at the
points of quasi-energy crossings.

\begin{figure}[htb]
\centering{\includegraphics[width=0.48\textwidth] {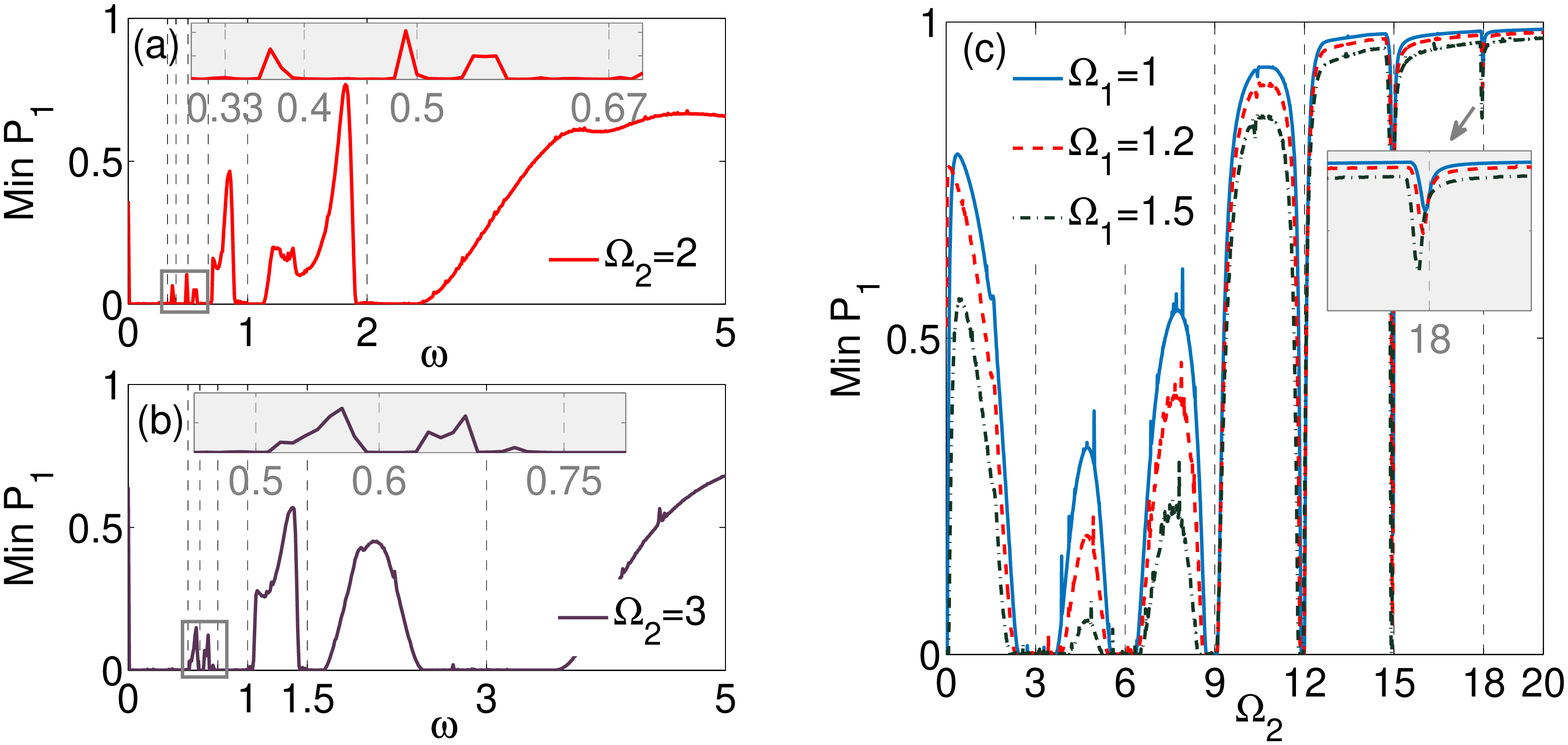}} \caption{(Color
online) (a)-(b)The minimum value of population distribution $ P_1 $,
$ \textrm{Min} (P_1) $,  as a function of the modulation frequency $
\omega $ with $ \Omega_2=2 $ and $ \Omega_2=3 $ respectively. Other
parameters are chosen as $ \Omega_1=1 $ and $ A=6.6$. (c)$
\textrm{Min} (P_1) $ versus $ \Omega_2 $ with different values of $
\Omega_1 $. Other parameters are chosen as $ \omega=3 $ and $ A=6.6
$. } \label{fig:fig3}
\end{figure}

In order to observe the $n$-photon-like resonances from a different
angle, we also plot $ \textrm{Min} (P_1) $ as a function of the
modulation frequency $ \omega $ for two fixed parameters $
\Omega_2=2 $ and $ \Omega_2=3 $ in Fig.\ref{fig:fig3} (a) and (b)
respectively. We can readily observe that the $n$-photon-like
resonances occur at comparatively broad intervals around
$\omega=\Omega_2/n$. The width of such photon-assisted tunneling
resonances is much larger than those of PAT resonances observed in
the literature. We now elaborate the physics underlying this
photon-assisted tunneling resonances. The actual resonance condition
does not refer directly to $\Omega_2=n\omega$ but rather to the
tunneling frequency of the model (\ref{equ:P3}) without periodic
modulation. The unmodulated three-guide optical system
(\ref{equ:P3}) admits three energy level as $0,\pm
\sqrt{\Omega_1^2+\Omega_2^2}$, and the space of two neighboring
energy levels is $\omega_0=\sqrt{\Omega_1^2+\Omega_2^2}$.  In such a
system, the existence of imbalanced dark state with zero energy
results in the suppression of tunneling when the periodic modulation
is switched off. The periodic modulation effectively creates
``photons" that bridge the energy gap between neighboring energy
levels. Thus, a photon-assisted tunneling resonance can occur at a
modulation frequency which satisfies the resonance condition
$\omega_0=n\omega$. When $\Omega_2$ is considerably larger than
$\Omega_1$, the energy difference $\omega_0$ of the unmodulated
system will become principally characterized by $\Omega_2$ and
therefore the resonance condition is approximately given by
$\Omega_2=n\omega$. As clearly seen in the inset in
Fig.\ref{fig:fig3} (c), the position of $n$-photon-like resonance
does slightly shift with increasing $\Omega_1$ due to the dependence
of the energy difference (tunneling frequency) $\omega_0$ on
$\Omega_1$.

\textbf{B. PAT in five-guide system}

We now turn to the case of the five-guide optical system where the
dynamical equations are

\begin{eqnarray}  \label{equ:P5}
i\frac{{da_1 }}{{dz}} &=& A \sin \left( {\omega z} \right)a_1 +
\Omega _1 a_2,
\nonumber \\
i\frac{{da_j }}{{dz}} &=& \Omega _1 a_{j+1} + \Omega _1 a_{j-1},~~~j=2,3   \nonumber \\
i\frac{{da_4 }}{{dz}} &= &\Omega_1 a_3+\Omega _2 a_5,  \nonumber \\
i\frac{{da_5 }}{{dz}} &= &\Omega _2 a_4.
\end{eqnarray}

\begin{figure}[htb]
\centering{\includegraphics[width=0.48\textwidth] {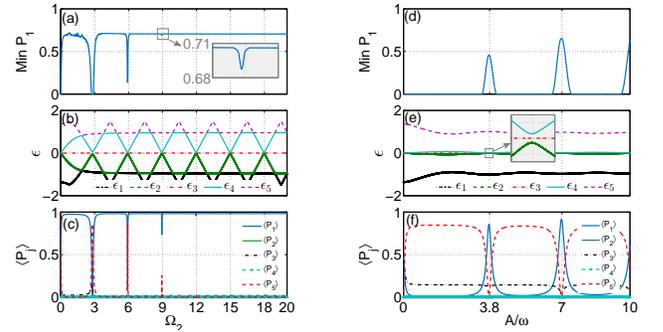}} \caption{(Color
online) PAT in five-guide optical system. The left column: (a) the
minimum value of population distribution at guide $1$, $ \textrm{Min
}(P_1) $, versus $ \Omega_2 $ with $ A=6.6, \omega=3,\Omega_1=1 $;
(b) the corresponding quasi-energy $ \epsilon $ versus $ \Omega_2 $
and (c)the time-averaged population $ \langle P_j \rangle $
belonging to the zero-energy (dark) Floquet state versus $\Omega_2$. The
right column: (d) $ \textrm{Min} (P_1) $ versus $ A/\omega $ for the
$1$-photon resonance $\omega=3, \Omega_2=2.8, \Omega_1=1$; (e) the
corresponding quasi-energy $ \epsilon $ versus $ A/\omega $ and (f)
the time-averaged population $ \langle P_j \rangle $ belonging to
the zero-energy (dark) Floquet state versus $A/\omega $.} \label{fig:fig4}
\end{figure}

The dynamics for the five-guide system are investigated by direct
integration of the time-dependent Schr\"{o}dinger equation
(\ref{equ:P5}) with the light initially localized at the guide $1$.
In Fig.\ref{fig:fig4}, we show the value of $ \textrm{Min} (P_1) $
as a function of $\Omega_2$ which exhibits a sequence of PAT
resonances with similar behavior as that of a three-guide system.
The higher $n$-photon resonances with $n\geq 3$ become very weak,
almost invisible, as illustrated in the inset of Fig.\ref{fig:fig4}
(a). By comparison of Fig.\ref{fig:fig2} (a) with Fig.\ref{fig:fig4}
(a), it is apparent that the same order PAT resonance for the
five-guide system is much narrower and weaker than for the
three-guide system. Like the case of the three-guide system, this
five-guide system also possesses a dark Floquet state with zero
quasi-energy and negligible population at all of the even $j$-th
guides (modes), as illustrated in Figs.\ref{fig:fig4}(b) and (c).
Reason for the existence of the analog of PAT resonances in the
five-guide system lies in that population distribution $\langle P_1
\rangle$ for the dark Floquet state simultaneously displays a series
of sharp dips at the positions of PAT resonances [see
Figs.\ref{fig:fig4}(c)].

In Fig.\ref{fig:fig4}(d), we plot $ \textrm{Min}(P_1) $ obtained in
the five-guide system as a function of the modulation parameter
$ A/\omega $ for the $1$-photon resonance $ \Omega_2=2.8, \omega=3,
\Omega_2\approx \omega $. As discussed before, we can observe that
the values of $ \textrm{Min}(P_1) $ are peaked at the zeros of
$J_1(A/\omega)$, at which CDT occurs, while between the peaks $
\textrm{Min}(P_1) $ take extremely low values as result of PAT.
However, compared with the case of three-guide system, the peaks in
$ \textrm{Min}(P_1)$ are considerably lower and broader. As can be
clearly seen from Figs.\ref{fig:fig4}(e) and (f), the peaks in $
\textrm{Min}(P_1) $ are indeed centered at the points of closest
approach of the quasi-energies where the dark Floquet state has a
population $\langle P_1\rangle>0.5$. The numerical results establish
again a firm link between PAT and dark Floquet state in our
considered systems.

\textbf{C. Tunneling dynamics in the four- and six-guide optical systems and beyond}

\begin{figure}[htb]
\centering{\includegraphics[width=0.48\textwidth] {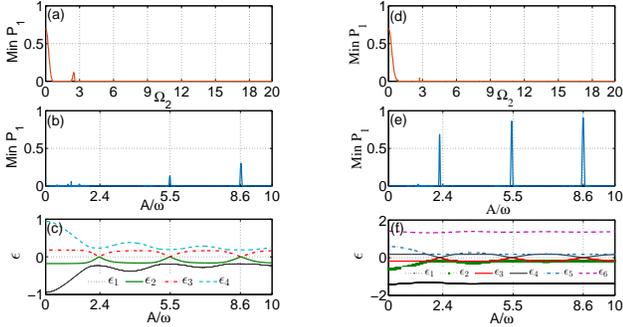}} \caption{(Color
online) The left column: the characteristics of four-guide optical
system; The right column: the characteristics of six-guide optical
system. (a) and (d): $\textrm{Min}(P_1)$ versus $ \Omega_2 $ at $
A=6.6, \omega=3, \Omega_1=1$; (b) and (e): $ \textrm{Min}(P_1) $
versus $A/\omega $ at $\Omega_2=3, \omega=3, \Omega_1=1$; (c) and
(f):
 quasi-energies $\varepsilon$ versus $ A/\omega $ at $\Omega_2=3, \omega=3,
\Omega_1=1$. }\label{fig:fig5}
\end{figure}

Finally, we briefly discuss the case of the four- and six-guide
systems. The dynamics for $ N=4 $ and $ N=6 $ are presented in
Fig.(\ref{fig:fig5}) on the basis of a full numerical analysis of
equation (\ref{equ:PN}) with the light initially populated in the
guide $1$. It tells the existence of a sharp transition from CDT to
complete tunneling for both cases of $N=4$ and $N=6$ when the
coupling strength $\Omega_2$ is increased from zero. A close
examination of the tunneling dynamics at $\Omega_2=n\omega$ shows
that the value of $\textrm{Min}(P_1)$ displays narrow peaks nearly
at zeros of $J_0(A/\omega)$ where a pair of quasi-energies become
degenerate. This closely resembles the case of the high-frequency
modulation $\omega\gg \textrm{max}(\Omega_1,\Omega_2)$ where CDT is
dominated by the zeros of $J_0(A/\omega)$. As shown in
Figs.(\ref{fig:fig5}) (b) and (e), the localization centered nearly
at zeros of $J_0(A/\omega)$ is fairly smaller for the four-guide
system, but still generates high peaks for the six-guide system.

Moreover, we have simulated multiwaveguide systems of other numbers
of waveguides. The numerical results, which are not displayed here,
show that the PAT resonances probably occur in all the odd-$N$-guide
optical systems, while the PAT resonances become weaker with the
increase of number of guides.  However, all the even-$N$-guide
optical systems exhibit CDT to complete tunneling transition without
appearance of $n$-photon-like resonance when the coupling strength
$\Omega_2$ is increased from zero, which is totally different from
the case of odd-$N$-guide system.

\section{Possibility of experimental realization }

Now, we discuss the experimental possibility of observing our
theoretical predictions based on the coupled-mode equations. A more
rigorous dynamics for our system can be simulated by the
Schr\"{o}dinger equation for the dimensionless field amplitude $ E
$, which describes the light propagation along the $ z $ axis of an
array of $ N $ waveguides

\begin{equation}  \label{equ:E1}
i\frac{\partial E}{\partial z}=-\frac{1}{2}\frac{\partial
^{2}E}{\partial ^{2}x}-pR(x,z)E.
\end{equation}

Here $ x $ and $ z $ are the normalized transverse and longitudinal
coordinates, and $ p $ describes the refractive index contrast of the
individual waveguide. For our system, the refractive index of the
first waveguide is harmonically  modulated along the propagation
direction, while all other $ N-1 $ waveguides are unmodulated. The
corresponding refractive index distribution of this kind of
waveguide system is given by

\begin{eqnarray}
R(x,z)&=&\sum_{j=-(N-2)}^{1}[1+f_{j}(z)]\exp\left[-\left(\frac{x-jw_{j}}{w_x}\right)^6\right],\\
f_{1}(z)&=&\mu\sin\left(\omega z\right),~~~f_{j}(z)=0 (j\neq
1),\nonumber
\end{eqnarray}

with the position of each waveguide being $ jw_j $, the channel width
$ w_x $, the longitudinal modulation amplitude $ \mu $, and the
modulation frequency $ \omega $. Therein the super-Gaussian function
$ \exp(-x^6/w_x^6) $ describes the profile of a single waveguide with
width $ w_x $. In our discussion, all the waveguide spacings
[$jw_j-(j-1)w_{j-1}$] are identical except that the spacing between
the bottom boundary waveguide and its neighbor is variable.

In what follows, we will illustrate our main results with a triplet
waveguide system ($ N=3 $) by directly integrating the field
propagation equation (\ref{equ:E1}) with realistic experimental
parameters. We set $ w_x=0.3, p=2.78, \mu=0.2 $ and
$ \omega=3.45\times(\pi/100) $. We characterize two distinct waveguide
spacings as $ ws_1 $ and $ ws_2 $ respectively, where $ ws_1=w_1 $
stands for the separation between the top waveguide and the middle
waveguide and $ ws_2=w_{-1} $ the separation between the bottom
waveguide and the middle waveguide. Further we set $ ws_1=3.2 $ and
choose different values of $ ws_2 $ to observe PAT resonance. As in
the experiments\cite{Szameit,Szameit2}, $ w_x $ and $ w_j $ are in units
of 10 $ \mu $m, and $ p = 2.78 $ corresponds to a real refractive index
of $ 3.1\times 10^{-4} $. In all simulations we excited the top
channel at $ z=0 $, using the fundamental linear mode of the isolated
waveguide. It is instructive to normalize the modulation frequency
to the beating frequency of the unmodulated linear dual-core coupler
with spacing $ ws_1 $, $ \Omega_b=2\Omega_1 = 2\pi/T_b $, where $ T_b $
is a beating period representing the shortest distance for the light
returning to the input waveguide. For our set of parameters one has
$ T_b =100 $ and thus $ \omega=3.45\Omega_1 $.

\begin{figure}[htb]
\centering{\includegraphics[width=0.48\textwidth] {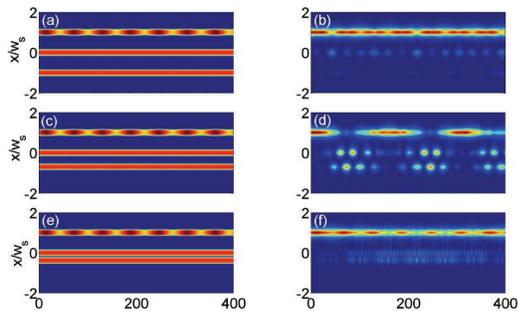}} \caption{(Color
online) Light propagation in three-guide optical systems with
different spacings between the bottom and the middle waveguide for
the input beam centered at the top waveguide. First row: (a) and
(b): the refractive index distribution $ R(x,z) $ and the light
propagation $ |E(x,z)|^2 $ for a three-guide system with equal
channel spacing $ ws_1=ws_2=3.2 $; Second row: (c) and
(d): the refractive index distribution $ R(x,z) $ and the light
propagation $ |E(x,z)|^2 $ for a three-guide system with unequal
channel spacing $ ws_1=3.2, ws_2=2.22 $; Third row: (e)
and (f): the refractive index distribution $ R(x,z) $ and the light
propagation $ |E(x,z)|^2 $ for a three-guide system with unequal
channel spacing $ ws_1=3.2, ws_2=1.2 $. }
\label{fig:fig6}
\end{figure}

The beam dynamics of a three-guide optical system are visualized in
Fig.(\ref{fig:fig6}) for three values of $ ws_2 $, which firmly
verifies the predictions from the coupled-mode equation
(\ref{equ:P3}). It can be readily observed from
Figs.(\ref{fig:fig6}) (a) and (b) that the light tunneling is almost
completely suppressed, as the three-channel waveguide system has
equal channel spacing $ ws_1=ws_2=3.2 $. At $ ws_2=2.22 $, the
light coupling between the waveguide channels is restored [see
Fig.(\ref{fig:fig6}) (c) and (d)]. The revival of light tunneling is
a signature of PAT resonance predicted by the coupled-mode theory.
In fact, our numerical simulation (not shown here) reveals that the
beating period of an unmodulated linear dual-core coupler with a
channel spacing $ 2.22 $ is about $ 100/3.45 $. As such, we have
$ \Omega_2\approx3.45\Omega_1 $ and $ \Omega_2\approx\omega $, which is
in fact the position of the first photon resonance. As the channel
spacing $ ws_2 $ is reduced further, it is  expectable to observe
again the strong suppression of light tunneling[see
Fig.(\ref{fig:fig6}) (e) and (f)]. These results are in good
qualitative agreement with those in Fig.\ref{fig:fig2}(a) based on
the coupled-mode equation.

\section{Conclusion}

 In summary, we have theoretically reported an analog of PAT in a
three-channel waveguide system, in which the space separation
between the bottom and the middle waveguides is adjustable and the
refractive index of the top waveguide is modulated periodically
along the light propagation direction. With the standard
coupled-mode theory, the system can be described by a driven
three-state discrete model with two distinct coupling strengths
$\Omega_1$ and $\Omega_2$, where $\Omega_1$ stands for the coupling
strength between states $1$ and $2$, and $\Omega_2$
 between states $2$ and $3$. In studying the three-state discrete model, we have found that (i) a strong
suppression (CDT) associated with the zero-energy (dark) Floquet
state persists even in the non-high-frequency modulation regimes
where $\omega\leq\textrm{max}(\Omega_1,\Omega_2)$ except at a series
of resonance positions; (ii) at particular values of the coupling
strength $\Omega_2$, $\Omega_2\approx
 n\omega$ with $n$ being integer, the tunneling dynamics is (partly) restored, analogous to the $n$-photon-like resonances which
 overcome the effect of suppression of tunneling.
The numerical calculations illustrate that the PAT resonances exist
in the five-state system and also probably exist in the systems with
arbitrary odd number of coupled states. In particular, the PAT
resonances will become
 weaker with the increase of number of states (modes). This type of
 PAT found in our work has a different origin from traditional PAT. It is closely related to the existence of the dark Floquet
 state. The main results are
demonstrated by the direct numerical simulations of propagation
dynamics based on the full continuous model with realistic
experimental parameters, which indicates that the PAT found in our
work can be readily tested in the current experimental setup.

\acknowledgments

The work is supported in part by the National Fundamental Research
Program of China (Grant No. 2012CB922103), the National Science
Foundation (NSF) of China (Grant Nos. 11375067, 11275074, 11374116,
11204096, 11405061 and 11574104), the Fundamental Research Funds for
the Central Universities, HUST (Grant No. 2014QN193). X. Luo is also
partially supported by the NSF of China under Grants 11465009,
11165009, the Program for New Century Excellent Talents in
University of Ministry of Education of China (NCET-13-0836),
Scientific and Technological Research Fund of Jiangxi Provincial
Education Department under Grant No. GJJ14566, and the financial
support provided by the Key Subject of Atomic and Molecular Physics
in Jiangxi Province. X. Luo thanks Prof. Biao Wu for his supports
over the years.

\end{document}